\documentclass{WileyMSP-template}

\usepackage{graphicx}

\usepackage[usenames,dvipsnames]{xcolor}

\newcommand{\rf}[1]{(\ref{#1})}

\newcommand{\beq}{\begin{equation}}
\newcommand{\eeq}{\end{equation}}
\newcommand{\beqr}{\begin{eqnarray}}
\newcommand{\eeqr}{\end{eqnarray}}
\newcommand{\lb}[1]{\label{#1}}
\newcommand{\bc}{\begin{center}}
\newcommand{\ec}{\end{center}}
\newcommand{\ct}[1]{\cite{#1}}

\begin{document}

\pagestyle{fancy}

\title{Oscillator laser model}

\maketitle


\author{Igor E. Protsenko*}
\author{Alexander V. Uskov}



\begin{affiliations}
I. E. Protsenko, A. V. Uskov\\
 P.N.Lebedev Physical Institute of the RAS,
Moscow 119991, Russia\\
Email Address: procenkoie@lebedev.ru, protsenk@gmail.com


\end{affiliations}



\begin{abstract}

A laser model is formulated in terms of quantum harmonic oscillators. Emitters in the low lasing  states are usual harmonic oscillators,  and emitters in the upper  states are {\em inverted} harmonic oscillators.  Diffusion coefficients, consistent with the model   and necessary for solving quantum nonlinear laser equations analytically, are found. Photon number fluctuations of the  lasing mode and fluctuations of  the population of the lasing states  are calculated. Collective Rabi splitting peaks are predicted in the intensity fluctuation spectra of the superradiant  lasers. Population fluctuation mechanisms in superradiant lasers and 
lasers without superradiance are discussed and compared with each other. 

\end{abstract}
\section{\label{Sec1}Introduction}
Recent technological developments  have led to a variety of novel miniature lasers  as quantum dot photonic crystal~\ct{7907235,Noda260,Prieto:15,Takiguchi:16,Ota:17,Nozaki:07,Yu2017}, 
micropillar~\ct{Li2019,doi:10.1063/1.4791563,Kreinberg2017}, plasmonic~\ct{Suh2012} and other kinds of nanolasers \ct{Khajavikhan2012}. All of them are intensively investigated. The motivation  for research of nanolasers is related  to fundamental questions,  such as general properties  of quantum fields with only a few photons in the mode and with practical applications,  such as the  direct incorporation of lasers  into nano-chips~\ct{Kurosaka2010,8658143,Crosnier2017}. Nano\-lasers often demonstrate  
 exciting and unusual properties,  such as the  mode-locking with a repetition rate independent of the cavity
length \ct{PhysRevA.102.043503,PhysRevLett.123.233901}.  

There is a  strong interest in {\it superradiant} (SR) lasers and nanolasers.  Such lasers  combine a  high gain with a small  cavity and operate in the  bad-cavity regime~\ct{Khanin, Belyanin_1998,PhysRevA.71.053818}. The active medium polarization is a dynamical variable in such lasers, where a collective spontaneous emission into the lasing mode is significant.  
SR lasers  have been experimentally realized  with cold alkaline earth atoms~\ct{PhysRevX.6.011025, PhysRevA.96.013847,PhysRevA.81.033847,PhysRevA.98.063837},  rubidium atoms~\ct{Bohnet}, and quantum dots~\ct{Jahnke} as the active medium. SR lasers are less sensitive  to the cavity-length fluctuations important for atomic clocks~\ct{PhysRevX.6.011025,PhysRevA.96.013847,Bohnet}. Superradiance leads to  interesting collective effects, such as excitation trapping~\ct{PhysRevA.81.033847,Bohnet} and superthermal photon statistics~\ct{Kreinberg2017,Jahnke,corr}, with possible applications in  high-visibility optical imaging~\ct{PhysRevA.95.053809}. 

Analytical description of SR lasers  meets difficulties. Quantum noise in such lasers is not a perturbation, the laser equations are nonlinear, and the active medium polarization is  a dynamical variable not eliminated adiabatically. Strong population fluctuations in SR lasers play a  significant role in laser dynamics leading to large relaxation oscillations \ct{Protsenko_2021} and the acceleration of spontaneous emission into the lasing mode \ct{Protsenko_2022}. High population fluctuations at superradiance were noted  previously, for example, at a generation of superradiant pulses \ct{PhysRevA.13.357}.

The first
purpose of  the present  work is 
to continue calculations and  investigation of the population fluctuations in SR lasers at  a low excitation  using the  approach of \ct{Protsenko_2021,Protsenko_2022,PhysRevA.59.1667,Andre:19}. The laser at a low excitation  does not generate coherent radiation and operates as a LED or, if the collective spontaneous emission (the superradiance)  into the lasing mode is significant, as SR LED. 

Calculations and investigation of the population fluctuations are necessary for better understanding properties of SR LEDs and lasers, for solving nonlinear laser equations \ct{Protsenko_2022}, and for finding quantum characteristics of  nanolasers as, for example, the  autocorrelation function $g_2$ of the lasing field,  widely used in experiments and the theory \ct{QO_book}. 

 Our previous papers \ct{PhysRevA.59.1667,Andre:19}, neglected population fluctuations at low laser excitation. In \ct{Protsenko_2021}, population fluctuations have been taken into account  at a high excitation when the laser generates coherent radiation  then laser equations can be solved by the usual perturbation procedure. The procedure of \ct{Protsenko_2021}  cannot account  for population fluctuations at a low laser excitation.   The paper \ct{Protsenko_2022} describes the treatment of nonlinearities and   population fluctuations at a low laser excitation. However, the analysis in  \ct{Protsenko_2022} was simplified and restricted by a very low laser excitation when population fluctuations determined by only the pump and the decay of the upper lasing states but practically do not depend on the lasing field. 

The restriction of analysis in \ct{Protsenko_2022} appears because of the difficulty in finding diffusion coefficients consistent with the approximate equations. It turns  out that the use of well-known diffusion coefficients found from generalized Einstein relations (GER) leads to inconsistencies in the results of the approximate equations, for example, the breaking of Bose-commutation relations for the lasing field operator \ct{Protsenko_2021}.
The fact that different approximate approaches in quantum optics require different diffusion coefficients is well-known. For example, some diffusion coefficients change in the transition from operator to c-number Langevin equations  -- see the discussion in section 12.3   in \ct{Scully}. 
  
Another purpose of this paper, necessary for  continuing  research  \ct{Protsenko_2022}, is to find   diffusion coefficients consistent with our approximate equations.  It will be done  with the help of an {\em oscillator laser model} (OLM), representing the laser as a combination of the usual and {\em inverted} harmonic oscillators. The model of optical media as a set of oscillators  is used widely in nonlinear optics \ct{PhysRevA.49.2065, DEMETER20131203}. Glauber introduced inverted oscillators  for the  modelling pump bath \ct{Glauber}. Inverted oscillators  found applications in the quantum theory of linear amplifiers \ct{Stenholm_1986}.  There is a difference between the quantum theory of linear amplifiers \ct{Stenholm_1986} and OLM. OLM does not eliminate the active medium polarisation adiabatically, quantum theory of linear amplifiers does this. 

OLM leads to the  same quantum laser equations as in \ct{Protsenko_2021,Protsenko_2022,Andre:19}, finds correct diffusion coefficients in the frame of  the input-output theory  \ct{PhysRevA.46.2766,PhysRevA.30.1386} and preserves Bose commutation relations for the field operators.  The results of OLM let us find population fluctuations in the first-order approximation.

Using approximate analytical equations with appropriate diffusion coefficients, we find and investigate the photon number and the population fluctuations, their spectra, and variances; describe sources of population fluctuations and compare population fluctuations in the SR LEDs and the LEDs without SR.

Section~\ref{Sec2} is introductory. There we describe the exact and approximate two-level laser models. In subsection \ref{subs_A}, we describe the approximate approach and derive Eqs.~\rf{binary_eqs_fl_1} -- \rf{binary_eqs_fl_3} for fluctuations of binary operators. 

Section~\ref{SecIII} describes the oscillator laser model. 

Section~\ref{Sec2a} presents the calculation of diffusion coefficients for Eqs.~\rf{binary_eqs_fl_1} -- \rf{binary_eqs_fl_3} consistent with the approximation when population fluctuations are neglected. There we find the population fluctuation spectrum in the first-order approximation.

Section~\ref{Sec4} contains results about the population fluctuation spectra and variances; describes the influence of population fluctuations on the field intensity fluctuation spectra.  This section  discusses the mechanisms of population fluctuations and the comparison of population fluctuations in SR LEDs and LEDs without SR. 

 Concluding section~\ref{Sec5} summarises the results. 

For convenience, Table~\ref{table} shows definitions  of operators, variables,  and parameters  with references to equations where the operator or the parameter appears in the main text. Table~\ref{table} does not include fluctuation operators, denoted by the symbol $\delta$, and mean values marked by the same letter as the operator. For example, there are no mean values like  $\Sigma = \left<\hat{\Sigma}\right>$ and fluctuation operators like  $\delta\hat{\Sigma} = \hat{\Sigma} - \Sigma$.

\begin{table}
 \caption{Definitions of operators and parameters}
  \begin{tabular}[htbp]{@{}llllll@{}}
    \hline
    Symbol & Definition & Eq. & Symbol & Definition &  Eq.\\
    \hline
    
    $\hat{a}$  & lasing mode amplitude  & \rf{las_H} & $\hat{\sigma}_i^+$ ($\hat{\sigma}_i$)  & rising (lowing) operator & \rf{las_H}\\
      &    of i-th emitter & & &of i-th emitter\\
      $\hat{n}^e_i$ ($\hat{n}^g_i$) & operator of population of &\rf{cr_H} & $\hat{v}$ & active medium polarisation &\rf{v_N_def}, \rf{pol_op}\\ &  excited (ground) state of i-th emitter & & &operator\\
      $\hat{N}_e$ ($\hat{N}_g$)  & operator of population &\rf{v_N_def} & $N$ & mean population inversion &Sec.~\ref{Sec1}\\
      & of all excited (ground) states & & &\\
      $\hat{F}_{\alpha}$&Langevin force operator &\rf{MBE_1}, \rf{MBE_2}, & $\hat{\Sigma}$&field-polarization interaction&\rf{Sigma_def0}, \rf{Sigma_def}\\
      &$\alpha=\{a,v,N_e,N_g,...\}$&etc. & &energy operator\\
      $\hat{n}$ & photon number operator & \rf{binary_eqs0_1} & $\hat{D}$& dipole-dipole interaction energy&\rf{n_and_D}\\& & & & operator &\\ $\hat{b}_i$ ($\hat{c}_i$) & Bose-operator for i-th emitter & \rf{Ham0} & $\hat{b}$ ($\hat{c}$) &Bose-operator for $N_e$ ($N_g$) emitters & \rf{b_and_c} \\ & in the excited (ground) state & & & in the excited (ground) states & \\
      $\hat{a}_{in}$ & laser field bath Bose-operator & \rf{eqs_abc_1} & $\hat{b}_{in}$, $\hat{c}_{in}$  & polarisation baths Bose-operators & \rf{eqs_abc_2}, \rf{eqs_abc_3}\\
      $\hat{D}_f$ &  dipole-dipole interaction energy & \rf{Eq_for_D} &$\hat{\Gamma}$ & interaction with bath environments& \rf{las_H}, \rf{Ham0} \\ & in terms of Bose-operators & & & \\
$\hat{\alpha}(\omega)$ & Fourier-component operator & \rf{FT_0} & $\delta\hat{\alpha}(\omega)$ & Fourier-component operator for& \rf{FT0}\\
& for $\hat{\alpha}(t)$, $\alpha=\{a,v,...\}$ & & &  fluctuations $\delta\hat{\alpha}(t)$, $\alpha=\{ n,\Sigma,D,... \}$  & \\
      $\Omega$ & vacuum Rabi frequency & \rf{las_H} & $f_i$ & normalized coupling strength  & \rf{las_H} \\     
      & & & &of i-th emitter with lasing mode & \\
$f_{bi}$ ($f_{ci}$) & normalized coupling strength of i-th  & \rf{Ham0} & $f_{b}$ ($f_{c}$) & average normalised coupling strength & \rf{b_and_c}\\  
& emitter in excited (ground)  state & & & of emitter in excited (ground) state & \\
      $f$ & average emitter-field  & \rf{av_coup}, \rf{f_def} & $\gamma_{\perp}$ & polarisation decay rate & \rf{MBE_2}   \\ & coupling strength &
& & &  \\ 
 $N_0$ & total number of emitters & \rf{las_H} & $\kappa$ & lasing mode amplitude decay rate & \rf{MBE_1}\\
$\gamma_{\parallel}$ & upper level population & \rf{MBE_3} &  $P$ & upper level dimensionless pump  & \rf{MBE_3} \\ & decay rate & & & rate, normalized to $\gamma_{\parallel}$ &\\ $2D_{\alpha\beta}$ & diffusion coefficient for Langevin &  \rf{dif_coef_def} & $\gamma_P$ & population fluctuation decay rate & \rf{MBE_St3}\\ &   force & & &  & \\ 
$N_{th}$ & threshold population inversion & \rf{n_field} & $n(\omega)$ & laser field spectrum & \rf{F_Spectr} \\
& in semiclassical laser theory & & & & \\
$\omega_0$ & optical carrier frequency & \rf{las_H}, \rf{Ham0} & $\omega$ & deviation from $\omega_0$ in  $n(\omega)$; frequency& \rf{FT_0}   \\ &    & & &   of intensity, etc. fluctuations  & \rf{FT0} \\
$\omega_{opt}$ & optical field frequency & after \rf{n_field} &  &  &  \\
$(\hat{a}\delta {{{\hat{N}}}_{e}})_{\omega}$ & Fourier-component of $\hat{a}\delta {{{\hat{N}}}_{e}}$ & \rf{power_sp_nonlin} & $S_{aN_e}(\omega)$ & power spectrum of $\hat{a}\delta {{{\hat{N}}}_{e}}$ & \rf{power_sp_nonlin}, \rf{Sp_conv}\\
$\delta^2{n}(\omega)$ & photon number fluctuation spectrum& \rf{Ph_n_fluc}, \rf{ph_fl_sp}& $\delta^2{n}$ & photon number fluctuation dispersion & \rf{ph_n_var}\\
$\delta^2N_e(\omega)$ & population fluctuation spectrum & \rf{Pop_fl_sp} & $\delta^2N_e$ & population fluctuation dispersion & \rf{pop_disp}\\
$S_{\Sigma}(\omega)$ & auxiliary function & \rf{Pop_fl_sp},  \rf{delta_Sigma}& $S(\omega)$ &  auxiliary function & \rf{Pop_fl_sp}, \rf{S_expr}\\
\hline
  \end{tabular}\label{table}
\end{table}
\section{\label{Sec2}Quantum lwo-level laser  model}
\begin{figure}
\bc\includegraphics[width=10cm]{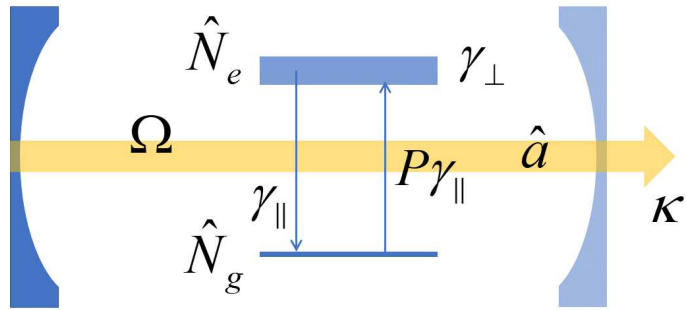}\ec
\caption{Scheme of the two-level laser. The population of the upper levels of emitters decays  to the low levels with the rate $\gamma_{\parallel}$   and  is pumped with the rate $\gamma_{\parallel}P$  from the low levels.  The lasing mode decays through the  semi-transparent mirror with the rate  $\kappa$ and resonantly interacts with lasing transitions of two-level emitters with the vacuum Rabi frequency $\Omega$. } 
\label{Fig01}
\end{figure}

We consider a lasing medium  with a large number $N_0 \gg 1$ of the two-level identical emitters in the optical cavity with the cavity mode resonant to the emitter transitions, shown schematically in Fig.~\ref{Fig01}. 
%
%
Emitters are fixed in space as in quantum dot lasers \ct{Jahnke,Eberl2000}. We investigate the stationary case when,  on average, $N_g$ emitters are in their ground states  and  $N_e$ emitters are in their excited states. 
The incoherent pump maintains the number of emitters in the excited states and the stationary population inversion $N=N_e-N_g$, compensating for the energy losses.

We consider a {\em weak excitation} of the lasing medium when:  (a)  the number of photons in the cavity is less or of the order of one and (b) there is no population inversion in the lasing medium. There is no generation of coherent radiation in such a regime, so the laser operates as a LED. We will call it a weak pump -- the pump providing the weak excitation regime.  

We write  Hamiltonian of the two-level laser, similar to the one used (maybe in  different notations) in many papers and books, as \ct{Protsenko_2021,Protsenko_2022, PhysRevA.59.1667,Andre:19, Scully,RevModPhys.68.127,trove.nla.gov.au/work/21304573}, written in the interaction picture with the carrier frequency  $\omega_0$ and the RWA approximation
\beq
        H=i\hbar \Omega \sum\limits_{i=1}^{{{N}_{0}}}{{{f}_{i}}\left( {{{\hat{a}}}^{+}}{{{\hat{\sigma }}}_{i}}-\hat{\sigma }_{i}^{+}\hat{a} \right)}+\hat{\Gamma }. \lb{las_H}
\eeq
Here $\Omega$  is the vacuum Rabi frequency; $f_i$  describes the difference in couplings of different emitters with the lasing mode, $\hat{a}$ is the Bose operator  of the lasing mode amplitude, 
$\hat{\sigma}_i$  is the lowing operator of i-th emitter \ct{RevModPhys.68.127}, $\hat{\Gamma}$  describes the interaction of the lasing mode and emitters with the white noise baths of the environment. 

Non-zero commutation relations for operators are \ct{Protsenko_2021,Protsenko_2022,PhysRevA.59.1667,Andre:19,RevModPhys.68.127}
\beq
        \left[ \hat{a},{{{\hat{a}}}^{+}} \right]=1,\hspace{0.25cm}  \left[ {{{\hat{\sigma }}}_{i}},\hat{\sigma }_{j}^{+} \right]=\left( \hat{n}_{i}^{g}-\hat{n}_{i}^{e} \right){{\delta }_{ij}},\hspace{0.25cm} \left[ {{{\hat{\sigma }}}_{i}},\hat{n}_{j}^{e} \right]=\left[ \hat{n}_{j}^{g},{{{\hat{\sigma }}}_{i}} \right]={{\delta }_{ij}}{{\hat{\sigma }}_{i}},\lb{cr_H}
\eeq
where $\hat{n}_{j}^{e}$  and $\hat{n}_{j}^{g}$  are operators of populations of the upper and the low levels of the i-th emitter, $\delta_{ij}$ is the Kronecker symbol. 

Following \ct{Protsenko_2021,Protsenko_2022,Andre:19}, we introduce operators  $\hat{v}$ and $\hat{N}_{e,g}$  of the polarization and  populations of all emitters 
\beq
\hat{v}=\sum\limits_{i=1}^{{{N}_{0}}}{{{f}_{i}}{{{\hat{\sigma }}}_{i}}},  \hspace{0.5cm} {{\hat{N}}_{e,g}}=\sum\limits_{i=1}^{{{N}_{0}}}{\hat{n}_{i}^{e,g}}. \lb{v_N_def}
\eeq
Using commutation relations \rf{cr_H} and Hamiltonian \rf{las_H}, we write Maxwell-Bloch equations for $\hat{a}$, $\hat{v}$, and ${\hat{N}}_{e}$ 
%
\beqr
  \dot{\hat{a}}&=&-\kappa \hat{a}+\Omega \hat{v}+{{{\hat{F}}}_{a}} \lb{MBE_1}\\
 \dot{\hat{v}}&=&-\left( {{\gamma }_{\bot }}/2 \right)\hat{v}+\Omega f\hat{a}\left( 2{{{\hat{N}}}_{e}}-{{N}_{0}} \right)+{{{\hat{F}}}_{v}} \lb{MBE_2}\\ 
 {{{\dot{\hat{N}}}}_{e}}&=&-\Omega \hat{\Sigma }+{{\gamma }_{\parallel }}\left[ P\left( {{N}_{0}}-{{{\hat{N}}}_{e}} \right)-{{{\hat{N}}}_{e}} \right]+{{{\hat{F}}}_{{{N}_{e}}}}, \lb{MBE_3} 
\eeqr
%
where 
\beq
\hat{\Sigma }={{\hat{a}}^{+}}\hat{v}+{{\hat{v}}^{+}}\hat{a}, \lb{Sigma_def0}
\eeq
$\kappa$, $\gamma_{\bot }$ and $\gamma_{\parallel }$  are decay rates,  $P\gamma_{\parallel }$ is the pump rate. We call dimensionless $P$ a normalized pump; $\hat{F}_{\alpha}$ with the index $\alpha =\left\{ a,v,N_e \right\}$  are Langevin forces of the white noise baths. The total number of emitters is preserved, so $\hat{N}_e+\hat{N}_g=N_0$.
In Eqs.~\rf{MBE_1} -- \rf{MBE_3}  we approximate 
\beq
f_{i}^{2}\approx f=N_{0}^{-1}\sum\limits_{i=1}^{{{N}_{0}}}{f_{i}^{2}}. \lb{av_coup}
\eeq
Physically, $\hat{\Sigma}$ is the operator of a normalized field-polarisation interaction energy 
analogous to the field-assisted polarisation  \ct{Jahnke,PhysRevA.75.013803}. An operator similar to $\hat{\Sigma}$ has been used previously, for example, in \ct{PhysRevA.62.043813}.

We rewrite Eqs.~\rf{MBE_1} -- \rf{MBE_3} in the form convenient for our calculations. We separate the operator mean values and fluctuations 
\beq
{{\hat{N}}_{e,g}}={{N}_{e,g}}+\delta{{\hat{N}}_{e,g}}, \hspace{0.5cm} \hat{\Sigma} = \Sigma + \delta\hat{\Sigma} \lb{fluc_mean}
\eeq
and write instead of Eqs.~\rf{MBE_1} -- \rf{MBE_3}
%
\beqr
  \dot{\hat{a}}&=&-\kappa \hat{a}+\Omega \hat{v}+{{{\hat{F}}}_{a}} \lb{MBE_St1} \\ 
 \dot{\hat{v}}&=&-\left( {{\gamma }_{\bot }}/2 \right)\hat{v}+\Omega f\left( \hat{a}N+2\hat{a}\delta {{{\hat{N}}}_{e}} \right)+{{{\hat{F}}}_{v}}. \lb{MBE_St2}\\
 {{{\delta\dot{\hat{N}}}}_{e}}&=&-\Omega \delta\hat{\Sigma }-\gamma_P\delta\hat{N}_e+\hat{F}_{N_e},  \lb{MBE_St3}
\eeqr
%
where $\gamma_P=\gamma_{\parallel }(P+1)$ and
\beq
        0=-\Omega \Sigma +{{\gamma }_{\parallel }}\left[ P\left( {{N}_{0}}-{{N}_{e}} \right)-{{N}_{e}} \right]. \lb{mean_pop}
\eeq
Diffusion coefficients for correlations of Langevin forces in Eq.~\rf{MBE_St1} -- \rf{MBE_St3}  are well-known \ct{RevModPhys.68.127,LandP} and derived from generalized Einstein relations  \ct{Lax_book1966}. 

We take the stationary mean photon number $n = \left<\hat{a}^+\hat{a}\right>$ and find from Eq.~\rf{MBE_St1}
$0=-2\kappa n +   \Omega\Sigma$.
The last equation and Eq.~\rf{mean_pop} lead to the energy conservation law
\beq
        2\kappa n=  \gamma_{\parallel}[P(N_0-N_e)-N_e]. \lb{eql_1}
\eeq
Nonlinear quantum equations \rf{MBE_St1} -- \rf{MBE_St3}, supplemented by Eqs.~\rf{Sigma_def0}, \rf{mean_pop}, and \rf{eql_1} are our  basic set of equations \ct{Protsenko_2021, Protsenko_2022,Andre:19}. This set is exact (in the frame of the two-level laser model), but it is hard to solve it analytically. We consider the stationary case and solve this set of equations by the approximate analytical approach described in the next subsection.
\subsection{Approximate analytical approach}\label{subs_A}
One can find an approximate analytical solution of  Eqs.~\rf{MBE_St1} -- \rf{MBE_St3} neglecting population fluctuations. We call this approach a zero-order approximation respectively to population fluctuations or simply a zero-order approximation. Taking $\delta{\hat{N}}_e=0$, we reduce the set \rf{MBE_St1} -- \rf{MBE_St3} to two linear, on $\hat{a}$ and $\hat{v}$, equations
%
\beqr
  \dot{\hat{a}}&=&-\kappa \hat{a}+\Omega \hat{v}+{{{\hat{F}}}_{a}} \lb{MBE_St01} \\ 
 \dot{\hat{v}}&=&-\left( {{\gamma }_{\bot }}/2 \right)\hat{v}+\Omega fN\hat{a}+\hat{F}_v. \lb{MBE_St02}
\eeqr
%
Non-zero diffusion coefficients $2D_{\alpha\beta}$ for Langevin forces in
\beq
\left<\hat{F}_{\alpha}(t)\hat{F}_{\beta}(t')\right> = 2D_{\alpha\beta}\delta{(t-t')} \lb{dif_coef_def}
\eeq
in Eqs.~\rf{MBE_St01}, \rf{MBE_St02}
are \ct{Protsenko_2021}:
\beq
2D_{aa^+} = 2\kappa, \hspace{0.5cm} 2D_{vv^+} = \gamma_{\perp}fN_g, \hspace{0.5cm}2D_{v^+v} = \gamma_{\perp}fN_e. \lb{dif_coef}
\eeq
Diffusion coefficients $2D_{vv^+}$ and $2D_{v^+v}$ are different from the "exact" ones found from GER when $\hat{N}_{e,g}$ are operators. We call diffusion coefficients \rf{dif_coef}, and the others found for the zero-order approximation equations as  zero-order diffusion coefficients. With  diffusion coefficients \rf{dif_coef},   Bose commutation relations for $\hat{a}$ found from  Eqs.~\rf{MBE_St01}, \rf{MBE_St02} are preserved.

One can solve linear Eqs.~\rf{MBE_St01}, \rf{MBE_St02} by the Fourier-transform
\beq
        \hat{\alpha}(t) = \frac{1}{\sqrt{2\pi}}\int_{-\infty}^{\infty}\hat{\alpha}(\omega)e^{-i\omega t}d\omega, \lb{FT_0}
\eeq
where $\alpha = \{a,v\}$ and calculate the field spectrum $n(\omega)$
\beq
\left<\hat{a}^+(\omega)\hat{a}(\omega')\right> = n(\omega)\delta(\omega+\omega'). \lb{F_Spectr}
\eeq
It was found in \ct{Protsenko_2021} that 
\beq
n(\omega)=\frac{(\kappa\gamma_{\perp}^2/2)N_{e}/N_{th}}{[(1-N/N_{th})(\kappa\gamma_{\perp}/2)-\omega^2]^2+\omega^2(\kappa+\gamma_{\perp}/2)^2}, \lb{n_field}
\eeq
where $N_{th} = 2\Omega^2f/\kappa\gamma_{\perp}$ is the threshold population inversion found in the semiclassical laser theory. The lasing field operator is $\hat{a}e^{-i\omega_0 t}$, so the frequency $\omega$ in Eqs.~\rf{FT_0} -- \rf{n_field}  is the {\em deviation} from the optical carrier frequency $\omega_0$. The field spectrum \rf{n_field}, expressed in terms  of the optical frequency $\omega_{opt} = \omega_0+\omega$, is $n(\omega_{opt}-\omega_0)$. Fig.~\ref{Fig1t} shows examples of the field spectra $n(\omega_{opt}-\omega_0)$.

With the help of Eq.~\rf{n_field}, we find the mean photon number $n=(2\pi)^{-1}\int_{-\infty}^{\infty}n(\omega)d\omega$
as a function of $N_e$. Then $N_e$ can be determined from the energy conservation law \rf{eql_1} as in \ct{Protsenko_2021,Protsenko_2022,PhysRevA.59.1667,Andre:19}.   

Zero-order approximation leads to interesting results, as the mean photon number $n(P)$ for  SR laser, found beyond the quantum rate equation approach of \ct{Coldren, doi:10.1063/1.5022958},  i.e. without adiabatic elimination of polarization; collective Rabi splitting in the laser field spectra \ct{Protsenko_2021,Protsenko_2022,Andre:19}.  Solving Eqs.~\rf{MBE_St01}, \rf{MBE_St02}  by Fourier transform, one can easily see that  the second-order autocorrelation function $g_2 \equiv \left<\hat{a}^+\hat{a}^+\hat{a}\hat{a}\right>/n^2=2$   in this zero-order approximation. One can suppose, therefore, that the super-thermal photon statistics with $g_2>2$ in the SR LEDs~\ct{Kreinberg2017,Jahnke,corr}  is due to  population fluctuations neglected in Eqs.~\rf{MBE_St01}, \rf{MBE_St02}.  

Recognizing the importance of population fluctuations in SR LEDs and lasers, we come in \ct{Protsenko_2022} to a first-order approximation, which includes population fluctuations. 

In the first-order approximation, we take into account the nonlinear term $\hat{a}\delta {{{\hat{N}}}_{e}}$ in Eq.~\rf{MBE_St2} and calculate $\delta {{{\hat{N}}}_{e}}$.  We consider the product $\hat{a}\delta {{{\hat{N}}}_{e}}$  as a  random variable with the Fourier-component operator $(\hat{a}\delta {{{\hat{N}}}_{e}})_{\omega}$ and the power spectrum $S_{aN_e}(\omega)$,
\beq
\left<(\hat{a}^+\delta {{{\hat{N}}}_{e}})_{\omega}(\hat{a}\delta {{{\hat{N}}}_{e}})_{\omega'}\right> = S_{aN_e}(\omega)\delta{(\omega+\omega')}. \lb{power_sp_nonlin}
\eeq
It was shown in \ct{Protsenko_2022} that $S_{aN_e}(\omega)$ is a convolution of the field spectrum $n(\omega)$ and the population fluctuation spectrum $\delta^2N_e(\omega)$
\beq
S_{aN_e}(\omega) = \frac{1}{2\pi}\int_{-\infty}^{\infty}n(\omega-\omega')\delta^2N_e(\omega')d\omega'. \lb{Sp_conv}
\eeq
A way for analytical calculation of $S_{aN_e}(\omega)$  is that $n(\omega)$ and $\delta^2N_e(\omega)$ in Eq.~\rf{Sp_conv} are found in the zero-order approximation.  So we calculate approximate $S_{aN_e}(\omega)$ from Eq.~\rf{Sp_conv} and then find the  first-order approximation solution of Eqs.~\rf{MBE_St1} -- \rf{MBE_St3} in terms of Fourier-component operators \ct{Protsenko_2022}.  

In \ct{Protsenko_2022}, we presented a simplified version of the first-order approximation,  supposing so weak excitation that the first term on the right in Eq.~\rf{MBE_St3} can be neglected, and   Eq.~\rf{MBE_St3} is approximately replaced by %
\beq
 \delta\dot{\hat{N}}_e\approx -\gamma_P\delta\hat{N}_e+\hat{F}_{N_e}.  \lb{red_pop_fl_eq}
\eeq
This equation can be easily solved.  

In \ct{Protsenko_2022}, we found that population fluctuations in SR LEDs significantly  enhance spontaneous emission into the lasing mode at certain conditions, increasing the LED radiation. This result confirms the importance of population fluctuations in SR LEDs and lasers.   It encourages us to continue working with the first-order approximation, mainly to carry it out without  Eq.~\rf{MBE_St3} replacement by Eq.~\rf{red_pop_fl_eq}. 
 
To find population fluctuations, in this case, we must know $\delta\hat{\Sigma}$. For this, we take Eqs.~\rf{MBE_St01}, \rf{MBE_St02}  and, applying the rule of the differentiation of products, write equations 
%
\beqr
  \dot{\hat{n}} &=&-2\kappa \hat{n}+\Omega \hat{\Sigma }+{{{\hat{F}}}_{n}} \lb{binary_eqs0_1}\\ 
 \dot{\hat{\Sigma }}&=&-\left( \kappa +{{\gamma }_{\bot }}/2 \right)\hat{\Sigma }+2\Omega f \left( \hat{n}N+\hat{D}+N_e \right)+{{{\hat{F}}}_{\Sigma }} \lb{binary_eqs0_2}\\
 \dot{\hat{D}}&=&-{{\gamma }_{\bot }}\hat{D}+\Omega N\hat{\Sigma }+\hat{F}_D, \lb{binary_eqs0_3} 
\eeqr
%
for the photon number operator $\hat{n} = \hat{a}^+\hat{a}$, $\hat{\Sigma}$, given by Eq.~\rf{Sigma_def0}, and
\beq
\hat{D} = f^{-1}\sum_{i\neq j}f_if_j\hat{\sigma}^{+}_i\hat{\sigma}_j, \lb{n_and_D} 
\eeq
which is  normalized dipole-dipole interaction energy of different emitters \ct{Protsenko_2021, Protsenko_2022,Andre:19}. $\hat{F}_{\alpha}$, $\alpha = \{n,\Sigma,D\}$ are Langevin forces.

We 
separate fluctuations and the mean values in operators
\beq
\hat{n} = n+\delta \hat{n}, \hspace{0.5cm} 
        \hat{\Sigma} = \Sigma+\delta \hat{\Sigma}, \hspace{0.5cm} \hat{D} = D+\delta\hat{D}, \lb{Fluc_Mean}
\eeq
insert Eqs.~\rf{Fluc_Mean} into Eqs.~\rf{binary_eqs0_1} -- \rf{binary_eqs0_3},  obtain equations for fluctuations
%
\beqr
  \delta\dot{\hat{n}} &=&-2\kappa \delta\hat{n}+\Omega \delta\hat{\Sigma }+{{{\hat{F}}}_{n}} \lb{binary_eqs_fl_1}\\ 
 \delta\dot{\hat{\Sigma }}&=&-\left( \kappa +{{\gamma }_{\bot }}/2 \right)\delta\hat{\Sigma }+2\Omega f \left( \delta\hat{n}N+\delta\hat{D} \right)+{{{\hat{F}}}_{\Sigma }} \lb{binary_eqs_fl_2}\\
 \delta\dot{\hat{D}}&=&-{{\gamma }_{\bot }}\delta\hat{D}+\Omega N\delta\hat{\Sigma }+\hat{F}_D \lb{binary_eqs_fl_3} 
\eeqr
%
and the set of algebraic equations for the mean values $n$, $\Sigma$ and $D$  
%
\beqr
  0 &=&-2\kappa n+\Omega \Sigma \lb{binary_eqs_st_1}\\ 
 0&=&-\left( \kappa +{{\gamma }_{\bot }}/2 \right)\Sigma+2\Omega f \left( nN+D+N_e \right) \lb{binary_eqs_st_2}\\
 0&=&-{{\gamma }_{\bot }}D+\Omega N\Sigma. \lb{binary_eqs_st_3} 
\eeqr
%
$n$, $\Sigma$  $D$ and $N_{e,g}$  found from  Eqs.~\rf{binary_eqs_st_1} -- \rf{binary_eqs_st_3} (with the energy conservation law \rf{eql_1} and the relation $N_e+N_g=N_0$)  are the same as the ones found from Eqs.~\rf{MBE_St1}, \rf{MBE_St2} 
\ct{Protsenko_2021,Andre:19}.

In the next steps of the procedure, we must solve linear Eqs.~\rf{binary_eqs_fl_1} -- \rf{binary_eqs_fl_3}  by the Fourier-transform, find $\delta\hat{\Sigma }$, find $\delta\hat{N}_e(\omega)$ from Eq.~\rf{MBE_St3} and calculate $\delta^2{N}_e(\omega)$ in Eq.~\rf{Sp_conv}. For doing this, we must know diffusion coefficients for Langevin forces in Eqs.~\rf{binary_eqs_fl_1} -- \rf{binary_eqs_fl_3}.

An essential part of our procedure is calculating the zero-order diffusion coefficients for Langevin forces in Eqs.~\rf{binary_eqs_fl_1} -- \rf{binary_eqs_fl_3}. In these equations (as well as in Eqs.~\rf{MBE_St01}, \rf{MBE_St02}), we can not  use "exact" diffusion coefficients found from GER; they will be inconsistent, as we will see, with the results of Eqs.~\rf{MBE_St01}, \rf{MBE_St02}.   

Calculation of the zero-order diffusion coefficients for Langevin forces in Eqs.~\rf{binary_eqs_fl_1} -- \rf{binary_eqs_fl_3} is carried out below in the frame of the oscillator laser model described in the next section. 
\section{Laser equations in terms of oscillators}\label{SecIII}
Two zero-order diffusion coefficients $2D_{vv^+}$ and $2D_{v^+v}$ for Langevin forces in Eqs.~\rf{MBE_St01}, \rf{MBE_St02} have been  determined in \ct{Protsenko_2022,Andre:19} from GER by setting populations to be constants (i.e. not operators).  We found it difficult to determine  {\em five} zero-order diffusion coefficients for Langevin forces in Eqs.~\rf{binary_eqs0_1} -- \rf{binary_eqs0_3} or \rf{binary_eqs_fl_1} -- \rf{binary_eqs_fl_3} similar way. It turns out to be easier to use a zero-order approximation {\em Hamiltonian} $H_0$ leading to Eqs.~\rf{binary_eqs0_1} -- \rf{binary_eqs0_3} or \rf{binary_eqs_fl_1} -- \rf{binary_eqs_fl_3}. The  properties of operators in $H_0$  help us to find  diffusion coefficients  for any equations of the zero-order approximation.  

\begin{figure}
\bc\includegraphics[width=10cm]{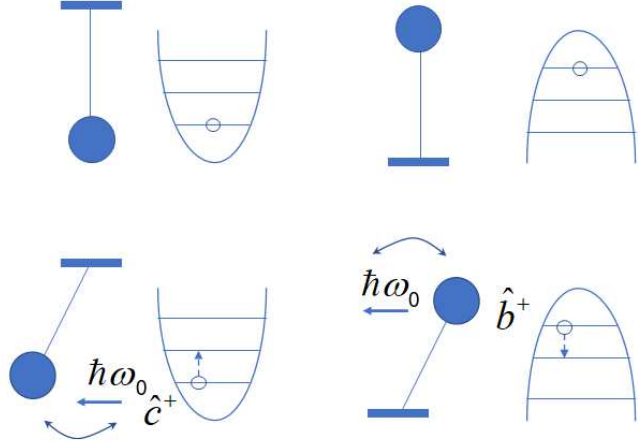}\ec
\caption{When the conventional  oscillator (on the left) accepts the energy quantum  $\hbar\omega_0$, it comes from the ground state to the first excited state. The inverted oscillator (on the right) loses a quantum and comes from its "inverted ground" state with the maximum energy to the first de-excited state. In the OLM, we consider virtual transitions, shown by arrows at the bottom of the figure and described by operators $\hat{b}^+$ for the inverted and $\hat{c}^+$ for the conventional  oscillator, see Hamiltonian \rf{Ham10}. We neglect changes in populations of levels and consider dynamics of the polarization of oscillators, see the discussion at the end of section \ref{Sec4}.} 
\label{Fig0t}
\end{figure}

Hamiltonian $H_0$ can be obtained in an {\em oscillator laser model} (OLM). 
This model describes  $N_g$ emitters in the ground states as conventional (normal) quantum harmonic oscillators with Bose operators $\hat{c}_ie^{-i\omega_0t}$, $i=1...N_g$. 
We will describe $N_e$ emitters in the upper  states  as {\em inverted} harmonic oscillators  with Bose-operators $\hat{b}_je^{+i\omega_0t}$, $j=1...N_e$.  Note  the sign "$+$" in the exponent multiplier $e^{+i\omega_0t}$ for the inverted oscillator, while the sign "$-$" in the multiplier $e^{-i\omega_0t}$ is for the usual, non-inverted oscillator.  $\hat{b}_j$ and $\hat{c}_j$ are  changed in time much slower than $e^{-i\omega_0t}$. The normal and the inverted oscillators  are shown schematically in Fig.~\ref{Fig0t}. 

%
%
%
Hamiltonian $H_0$ of the OLM, written in the interaction picture and the RWA with the carrier frequency $\omega_0$ is
\beq
H_0=i\hbar\Omega\left[ \hat{a}^+\left(\sum_{i=1}^{N_e}f_{bi}\hat{b}_i^+ + \sum_{i=1}^{N_g}f_{ci}\hat{c}_i\right) - h.c. \right]+\hat{\Gamma}. \lb{Ham0}
\eeq
Here 
dimensionless factors $f_{b,ci}$  describe the difference in couplings of different oscillators with the field; the rest of the notations is the same as for the "exact" two-level laser Hamiltonian \rf{las_H}. 

We represent $H_0$  in terms of only  three oscillators with Bose operators $\hat{a}$, $\hat{b}$ and $\hat{c}$
\beq
        \hat{b} = f_b^{-1}\sum_{i=1}^{N_e}f_{bi}\hat{b}_i, \hspace{0.5cm} \hat{c} =f_c^{-1}\sum_{i=1}^{N_g}f_{ci}\hat{c}_i, \hspace{0.5cm}  f_{b,c} = \left(\sum_{i=1}^{N_{e,g}}f_{b,ci}^2\right)^{1/2}.\lb{b_and_c}
\eeq
Bose commutation relations  $[\hat{b},\hat{b}^+] = [\hat{c},\hat{c}^+] = 1$  follow from Bose commutation relations for $\hat{b}_i$ and $\hat{c}_i$. Hamiltonian \rf{Ham0}, in terms of three oscillators, reads
\beq
H_0=i\hbar\Omega\left[ f_b(\hat{a}^+\hat{b}^+ - \hat{a}\hat{b}) + f_c(\hat{a}^+\hat{c} - \hat{c}^+\hat{a})\right]+\hat{\Gamma}. \lb{Ham10}
\eeq
Hamiltonian \rf{Ham10} and Bose-commutation relations  lead to Heisenberg-Langevin equations for $\hat{a}$, $\hat{b}^+$ and $\hat{c}$
%
\beqr
  \dot{\hat{a}}& = & -\kappa \hat{a}+\Omega \left( {{f}_{b}}{{{\hat{b}}}^{+}}+{{f}_{c}}\hat{c} \right)+\sqrt{2\kappa }{{{\hat{a}}}_{in}} \lb{eqs_abc_1}\\
 \dot{\hat{b}}^{+} & =& -\left( {{\gamma }_{\bot }}/2 \right){{{\hat{b}}}^{+}}+\Omega {{f}_{b}}\hat{a}+\sqrt{{{\gamma }_{\bot }}}{{{\hat{b}}}^+_{in}} \lb{eqs_abc_2}\\ 
  \dot{\hat{c}}&=&-\left( {{\gamma }_{\bot }}/2 \right)\hat{c}-\Omega {{f}_{c}}\hat{a}+\sqrt{{{\gamma }_{\bot }}}{{{\hat{c}}}_{in}},  \lb{eqs_abc_3}
\eeqr
%
where 
$\hat{a}_{in}$, $\hat{b}_{in}$ and $\hat{c}_{in}$ are  Bose-operators of baths.  Non-zero correlations between the bath operators are
\beq
        \left<\hat{\alpha}_{in}(t)\hat{\alpha}_{in}^+(t')\right> = \delta(t-t'), \lb{comm_Bose}
\eeq
where $\hat{\alpha}_{in} = \{\hat{a}_{in},\hat{b}_{in},\hat{c}_{in}\}$. Operators of different baths do not correlate with each other. Langevin forces with the bath operators are added in Eqs.~\rf{eqs_abc_1} -- \rf{eqs_abc_3} following the input-output theory  \ct{PhysRevA.46.2766,PhysRevA.30.1386}.

We simplify Eqs.~\rf{eqs_abc_1} -- \rf{eqs_abc_3} by  introducing an operator 
\beq
\hat{v}={{f}_{b}}{{\hat{b}}^{+}}+{{f}_{c}}\hat{c} \lb{pol_op}
\eeq
of the polarization of the lasing medium (using the same notation $\hat{v}$ as in Eq.~\rf{v_N_def}), and re-write Eqs.~\rf{eqs_abc_1} -- \rf{eqs_abc_3}
%
\beqr
   \dot{\hat{a}}&=&-\kappa \hat{a}+\Omega \hat{v}+\sqrt{2\kappa }{{{\hat{a}}}_{in}} \lb{eqs_z_od_1}\\ 
  \dot{\hat{v}}&=&-\left( {{\gamma }_{\bot }}/2 \right)\hat{v}+\Omega \left( f_{b}^{2}-f_{c}^{2} \right)\hat{a}+{{{\hat{F}}}_{v}}, \lb{eqs_z_od_2} 
\eeqr
%
with the Langevin force
\beq
{{\hat{F}}_{v}}=\sqrt{{{\gamma }_{\bot }}}\left( {{f}_{b}}{{{\hat{b}}}^+_{in}}+{{f}_{c}}{{{\hat{c}}}_{in}} \right).\lb{LF_5}
\eeq
Langevin forces are delta-correlated
$\left<\hat{F}_{\alpha}(t)\hat{F}_{\beta}(t')\right> = 2D_{\alpha\beta}\delta(t-t')$ 
with diffusion coefficients
\beq
2D_{v^+v} = \gamma_{\perp}f_b^2,\hspace{0,5cm} 2D_{vv^+} = \gamma_{\perp}f_c^2,\hspace{0,5cm}2D_{v^+v} = 2D_{vv^+} = 0, \lb{dif_coef_1}
\eeq
followed from Eqs.~\rf{comm_Bose}, \rf{pol_op} and \rf{LF_5}. We approximate %
\beq
f_{b,c}^2\approx fN_{e,g}, \hspace{0.5cm} f=N_0^{-1}\left(\sum_{i=1}^{N_{e}}f_{bi}^2+\sum_{i=1}^{N_{g}}f_{ci}^2\right), \lb{f_def}
\eeq
where ${{N}_{e}}$,  (${{N}_{g}}$) are the mean numbers of emitters in the ground (in the excited) states, and ${{N}_{0}}={{N}_{e}}+{{N}_{g}}$ is the total number of emitters. With the approximation \rf{f_def}, equations \rf{eqs_z_od_1}, \rf{eqs_z_od_2} became identical with equations \rf{MBE_St01}, \rf{MBE_St02} and diffusion coefficients \rf{dif_coef_1} -- with diffusion coefficients \rf{dif_coef}.
Such a coincidence 
 confirms the correctness of the OLM.
\section{\label{Sec2a} Diffusion coefficients and population fluctuations}
Now we will see that the OLM leads to Eqs.~\rf{binary_eqs_fl_1} -- \rf{binary_eqs_fl_3} and obtain diffusion coefficients for these equations.  
For this, using Eqs.~\rf{eqs_z_od_1}, \rf{eqs_z_od_2} and the  rule of differentiation of products, we write  the equation of motion for  $\hat{n}={{\hat{a}}^{+}}\hat{a}$ 
\beq
	\dot{\hat{n}}=-2\kappa \hat{n}+\Omega\hat{\Sigma }+{{\hat{F}}_{n}}, \lb{ph_n_eq}
\eeq
where
\beq
\hat{\Sigma }={{f}_{b}}\left( \hat{b}\hat{a}+{{{\hat{a}}}^{+}}{{{\hat{b}}}^{+}} \right)+{{f}_{c}}\left( {{{\hat{c}}}^{+}}\hat{a}+{{{\hat{a}}}^{+}}\hat{c} \right),\lb{Sigma_def}
\eeq
and the Langevin force
\beq
	{{\hat{F}}_{n}}=\sqrt{2\kappa }\left( {{{\hat{a}}}^+_{in}}\hat{a}+{{{\hat{a}}}^{+}}{{{\hat{a}}}_{in}} \right). \lb{LF_ph_n}
\eeq
In Eqs.~\rf{Sigma_def0}, \rf{Sigma_def},  and below,  we use the same notation $\Sigma$  for the normalized field-polarization interaction energy. We write, using Eqs.~\rf{eqs_z_od_1}, \rf{eqs_z_od_2},  
\beq\dot{\hat{\Sigma }}=-\left( \kappa +{{\gamma }_{\bot }}/2 \right)\hat{\Sigma }+2\Omega \left[ \hat{n}\left( f_{b}^{2}-f_{c}^{2} \right)+\hat{D}_f+f_{b}^{2} \right]+{{\hat{F}}_{\Sigma }}, \lb{Eq_sigma}
\eeq
where
\beq
\hat{D}_f=f_{b}^{2}{{\hat{b}}^{+}}\hat{b}+f_{c}^{2}{{\hat{c}}^{+}}\hat{c}+{{f}_{b}}{{f}_{c}}\left( {{{\hat{c}}}^{+}}{{{\hat{b}}}^{+}}+\hat{c}\hat{b} \right),\lb{D_def}
\eeq
and the Langevin force
\beqr
{{\hat{F}}_{\Sigma }}&=&\sqrt{2\kappa }{{f}_{b}}\left( \hat{b}{{{\hat{a}}}_{in}}+{{{\hat{b}}}^{+}}\hat{a}_{in}^{+} \right)+\sqrt{{{\gamma }_{\bot }}}{{f}_{b}}\left( {{{\hat{b}}}_{in}}\hat{a}+\hat{b}_{in}^{+}{{{\hat{a}}}^{+}} \right)+\nonumber\\& &\sqrt{2\kappa }{{f}_{c}}\left( {{{\hat{c}}}^{+}}{{{\hat{a}}}_{in}}+\hat{a}_{in}^{+}\hat{c} \right)+\sqrt{{{\gamma }_{\bot }}}{{f}_{c}}\left( \hat{c}_{in}^{+}\hat{a}+{{{\hat{a}}}^{+}}{{{\hat{c}}}_{in}} \right).\lb{F_sigma0}
\eeqr
With the derivation of Eq.~\rf{Eq_sigma}, we set the normal ordering of Bose operators replacing  $f_b^2\hat{a}\hat{a}^+$ with $f_b^2\hat{a}^+\hat{a}+f_b^2$.  
We obtain the equation for $\hat{D}_f$ 
\beq
\dot{\hat{D}}_f=-{{\gamma }_{\bot }}\hat{D}_f+\Omega \left( f_{b}^{2}-f_{c}^{2} \right)\hat{\Sigma }+{{\hat{F}}_{D_f}}\lb{Eq_for_D}
\eeq
with the Langevin force
\beqr
{{\hat{F}}_{D_f}}&=&\sqrt{{{\gamma }_{\bot }}}\left[ f_{b}^{2}\left( {{{\hat{b}}}^+_{in}}\hat{b}+{{{\hat{b}}}^{+}}{{{\hat{b}}}_{in}} \right)+f_{c}^{2}\left(
\hat{c}^+_{in}\hat{c}+{{{\hat{c}}}^{+}}{{{\hat{c}}}_{in}} \right)+\right.\nonumber\\&
&\left.{{f}_{b}}{{f}_{c}}\left( {{{\hat{c}}}^+_{in}}{{{\hat{b}}}^{+}}+{{{\hat{c}}}^{+}}{{{\hat{b}}}^+_{in}}+{{{\hat{c}}}_{in}}\hat{b}+\hat{c}{{{\hat{b}}}_{in}} \right) \right]\lb{LF_D0}
\eeqr
in the same way as equations for $\hat{n}$ and $\hat{\Sigma}$. 

Eqs.~\rf{ph_n_eq}, \rf{Eq_sigma} and \rf{Eq_for_D} are the set of linear equations for $\hat{n}$, $\hat{\Sigma}$ and $\hat{D}_f$. 
Delta-correlations for Langevin forces in these equations have non-zero diffusion coefficients
\beqr
&2{{D}_{nn}}=2\kappa n, \hspace{0.5cm} 2{{D}_{\Sigma \Sigma }}=2\kappa D_f+{{\gamma }_{\bot }}\left( f_{b}^{2}+f_{c}^{2} \right)n+\left( 2\kappa +{{\gamma }_{\bot }} \right)f_{b}^{2}&\nonumber\\
&2{{D}_{D_fD_f}}={{\gamma }_{\bot }}\left[ \left( f_{c}^{2}+f_{b}^{2} \right)D_f+2f_{b}^{2}f_{c}^{2} \right],&\lb{dif_n}\\
&2D_{\Sigma n} = 2D_{n\Sigma} = \kappa\Sigma, \hspace{0.5cm} 2D_{\Sigma D_f} = 2D_{D_f\Sigma} = (\gamma_{\perp}/2)(f_b^2+f_c^2)\Sigma.\nonumber&
\eeqr
We use approximation \rf{f_def}, introduce $D=fD_f$ and see Eqs.~\rf{ph_n_eq}, \rf{Eq_sigma} and \rf{Eq_for_D} equivalent to  Eqs.~\rf{binary_eqs0_1} -- \rf{binary_eqs0_3}. Separating the mean values and fluctuations in operators as in Eqs.~\rf{Fluc_Mean} we come to Eqs.~\rf{binary_eqs_fl_1} -- \rf{binary_eqs_fl_3}. Diffusion coefficients for Langevin forces in Eqs.~\rf{binary_eqs0_1} -- \rf{binary_eqs0_3} or \rf{binary_eqs_fl_1} -- \rf{binary_eqs_fl_3}  are 
\beqr
&2{{D}_{nn}}=2\kappa n, \hspace{0.5cm} 2{{D}_{\Sigma \Sigma }}=f[2\kappa D+{{\gamma }_{\bot }}N_0n+\left( 2\kappa +{{\gamma }_{\bot }} \right)N_e]&\nonumber\\
&2{{D}_{DD}}={{\gamma }_{\bot }}(N_0D+2N_eN_g),&\lb{dif_n0}\\
&2D_{\Sigma n} = 2D_{n\Sigma} = \kappa\Sigma, \hspace{0.5cm} 2D_{\Sigma D} = 2D_{D\Sigma} = (\gamma_{\perp}/2)N_0\Sigma.\nonumber&
\eeqr
Diffusion coefficients followed from GER are all the same as in Eqs.~\rf{dif_n0}, apart from $2D_{DD}$.  The term $\gamma_{\perp}2N_eN_g$  is absent in $2D_{DD}$  found in GER. 

We make the Fourier transform 
\beq
\delta\hat{\alpha}(t) = \frac{1}{\sqrt{2\pi}}\int_{-\infty}^{\infty}\delta\hat{\alpha}(\omega)e^{-i\omega t}d\omega,\hspace{0.5cm}\delta\hat{\alpha} = \{ \delta\hat{n},\delta\hat{\Sigma },\delta\hat{D} \} \lb{FT0}
\eeq
in the linear equations \rf{binary_eqs_fl_1} -- \rf{binary_eqs_fl_3} and obtain algebraic equations for Fourier-component operators.  
Solving them 
we find $\delta\hat{n}(\omega )$, $\delta \hat{\Sigma }(\omega )$.  From the relation
\beq
\left<\delta\hat{n}(\omega) \delta\hat{n}(\omega')\right>=\delta^2{n}(\omega )\delta(\omega
+\omega') \lb{Ph_n_fluc}
\eeq
and the similar one for $\delta \hat{\Sigma }(\omega )$, we find the photon number fluctuation    spectra $\delta^2n(\omega)$ and  $\delta^2\Sigma(\omega)$ with explicit expressions   presented in the Appendix. We use the notation $\omega$ for the oscillation frequency in Eq.~\rf{FT0} and everywhere below. It should not be confused with the same notation $\omega$  in Eqs.~\rf{FT_0}, \rf{F_Spectr}, and in Eq. \rf{n_field} for the field spectrum $n(\omega)$, where $\omega$ means the deviation from the optical carrier frequency $\omega_0$. 

We obtain the equation for the Fourier component operator $\delta\hat{N}_e(\omega)$ from the linear equation \rf{MBE_St3}. We insert there $\delta \hat{\Sigma }(\omega )$ found above, find $\delta\hat{N}_e(\omega )$ and, from the relation $\left<\delta\hat{N}_e(\omega)\delta\hat{N}_e(\omega')\right> = \delta^2{N}_e(\omega)\delta(\omega+\omega')$, obtain the population fluctuation spectrum 
\beq
        \delta^2{N}_e(\omega) = \frac{S_{\Sigma}(\omega )}{(\omega^2+\gamma_P^2)S(\omega )}+\frac{2D_{N_eN_e}}{\omega^2+\gamma_P^2}.\lb{Pop_fl_sp}
\eeq
Eq.~\rf{Pop_fl_sp}  is  the first-order approximation for $\delta^2{N}_e(\omega)$ since  $S_{\Sigma}(\omega )$ and $S(\omega )$, given in Appendix, are calculated in the zero-order approximation and do not depend on population fluctuations.  Writing Eq.~\rf{Pop_fl_sp},  we neglect quantum correlations between the population and the polarisation fluctuations of emitters, as we did in   \ct{Protsenko_2021,Protsenko_2022}. It is a good approximation if the total number of emitters is considerable so that the mean number of photons $n\ll N_{e,g}$.

We take the diffusion coefficient for population fluctuations 
\beq
2D_{N_eN_e}=\gamma_{\parallel}(PN_g+N_e) \lb{Dif_c_pop}
\eeq
the same as in \ct{Protsenko_2021}, the rate equation laser theory \ct{Coldren} and  calculations from GER \ct{Lax_book1966}.

We find that the photon number variance,  
\beq
\delta^2n = \frac{1}{2\pi}\int_{-\infty}^{\infty}\delta^2n(\omega)d\omega = n(n+1), \lb{ph_n_var}
\eeq
is the same as for the thermal radiation,  consistent with the second-order correlation function $g_2=2$ found from the solution of Eqs.~\rf{MBE_St01}, \rf{MBE_St02}. 
The result \rf{ph_n_var} confirms the correctness of the zero-order diffusion coefficients \rf{dif_n0}. 
One can not obtain the result \rf{ph_n_var} with diffusion coefficients from GER, i.e. without the term $\sim N_eN_g$ in $2D_{DD}$ in Eq.~\rf{dif_n0}. 
\section{\label{Sec4} Results and discussion}
We calculate zero-order diffusion coefficients \rf{dif_n0} and found population fluctuations  in the first-order approximation from Eqs.~\rf{binary_eqs_fl_1} -- \rf{binary_eqs_fl_3} and \rf{MBE_St3}. Diffusion coefficients \rf{dif_n0} let us continue the research of \ct{Protsenko_2022} and find the approximate analytical solution of Eqs.~\rf{MBE_St1} -- \rf{MBE_St3} without replacing Eq.~\rf{MBE_St3} with Eq.~\rf{red_pop_fl_eq}. We will find this solution in the future.

Below we investigate the photon number and  the   population fluctuations using their spectra and variances. We   compare the results for SR LEDs and LEDs without SR.

In examples we take the values of parameters the same as in \ct{Protsenko_2021,Protsenko_2022}, $\Omega/\gamma_{\parallel} = 34$, $N_0=100$,  $f=1/2$, $\kappa/\gamma_{\parallel} = 50$, where $\gamma_{\parallel}= 10^9$~rad/sec and $\kappa = 50\cdot 10^9$~~rad/sec. Similar parameter values can be found, for example, in  quantum dot lasers with photonic crystal cavities   \ct{doi:10.1063/1.5022958,PhysRevB.46.15574}.

We vary the polarisation relaxation rate $\gamma_{\perp}$ such that $1/15 < 2\kappa/\gamma_{\perp} < 2$. The case of $2\kappa/\gamma_{\perp} \geq 1$ corresponds to SR lasers,  $2\kappa/\gamma_{\perp}\ll 1$ - to lasers without SR   \ct{Protsenko_2021}.  
According to Figs.~3 and 4 of \ct{Protsenko_2021}, the laser at the chosen parameter values has a small mean cavity photon number  $n<1$ and operates in the LED regime for $P\leq 3$. So the weak laser excitation is in the interval  $0<P\leq 3$, and we vary the normalized pump rate $P$ in this interval. 
The values of all other laser parameters, apart from $P$ and $\gamma_{\perp}$, are fixed.

\begin{figure}
\bc\includegraphics[width=10cm]{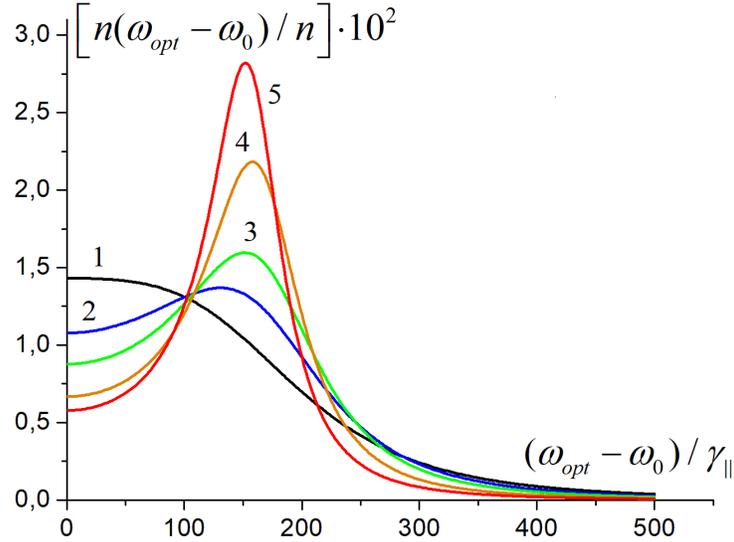}\ec
\caption{Laser field spectra as functions of the optical frequency $\omega_{opt}=\omega+\omega_0$ for $P=0.7$,  $2\kappa/\gamma_{\perp}  = 1/5$ (curve 1), $1/3$ (2), $1/2$ (3), $1$ (4) and $2$ (5), with other parameters, fixed and given in the text. Parts of spectra with $\omega_{opt}-\omega_0<0$ are symmetric with parts with $\omega_{opt}-\omega_0>0$ shown in the figure. All curves, but curve 1, have peaks of CRS near $(\omega_{opt}-\omega_0)/\gamma_{\parallel}\approx 150$. CRS peaks grow when the laser approaches the SR regime increasing $2\kappa/\gamma_{\perp}$.}
\label{Fig1t}
\end{figure}

Fig.~\ref{Fig1t} shows the field spectra $n(\omega_{opt}-\omega_0)$ given by Eq.~\rf{n_field}. We take $P=0.7$ and decrease $\gamma_{\perp}$ from  curve $1$ to curve $5$ in Fig.~\ref{Fig1t}, so   $2\kappa/\gamma_{\perp}  = 1/5$ (curve 1), $1/3$ (2), $1/2$ (3), $1$ (4), $2$ (5) and the contribution of SR in the laser grows from the curve 1 to the curve 5. In  Fig.\ref{Fig1t}, we  see peaks of the collective Rabi splitting (CRS) \ct{Andre:19} on curves 2 - 5 with  $2\kappa/\gamma_{\perp}  \geq 1/3$. The maxima of the CRS peaks grow with  $2\kappa/\gamma_{\perp}$, i.e. when the laser approaches the superradiant regime. 
%
%
%
%

\begin{figure}
\bc\includegraphics[width=10cm]{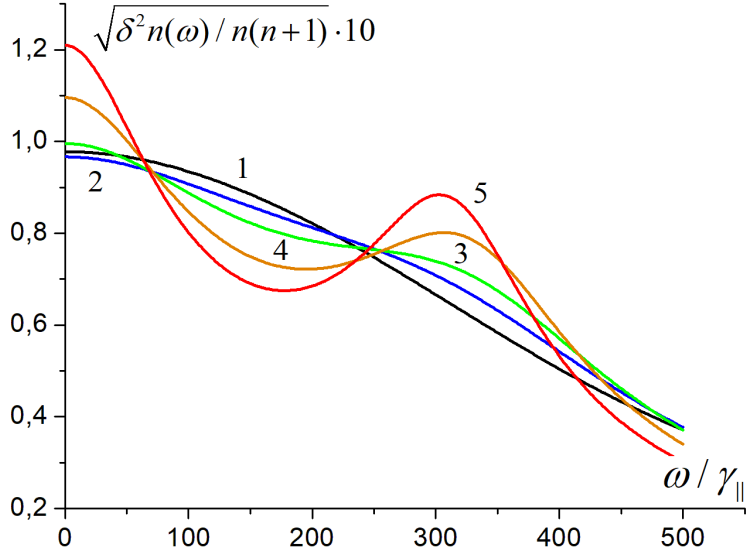}\ec
\caption{Photon number fluctuation spectra for various $\gamma_{\perp}$, the same as in Fig.~\ref{Fig1t} for curves  $1-5$ correspondingly, and other parameters  are given in the text. Curves 3, 4 and 5 have  peaks of CRS at frequencies $\omega$ approximately two times larger than the values of $\omega_{opt} - \omega_0$ at the maxima of the CRS peaks of the field spectra in Fig.~\ref{Fig1t}.}
\label{Fig2t}
\end{figure}

Fig.~\ref{Fig2t} shows the normalized photon number (or, the same, the field intensity) fluctuation spectra\\ $\sqrt{\delta^2n(\omega)/n(n+1)}$ for  the same parameter values as for Fig.~\ref{Fig1t},  $\delta^2n(\omega)$ is given by Eq.~\rf{ph_fl_sp} of Appendix. 

When $2\kappa/\gamma_{\perp}$ is relatively large, as for curves 3, 4 and 5 in Fig.\ref{Fig2t},  we see the sideband CRS peaks on curves.  The maxima of the CRS peaks in Fig.~\ref{Fig2t} correspond approximately to double the  
values of $\omega_{opt}-\omega_0$ of the CRS peak maxima in the field spectra of Fig.~\ref{Fig1t}. Formally, such a doubling is because $\hat{n} =\hat{a}^+\hat{a}$ is the binary operator in $\hat{a}$ and $\hat{a}^+$. The relative heights of the CRS peaks in the intensity fluctuation spectra in Fig.~\ref{Fig2t} are smaller than in the field spectra in  Fig.~\ref{Fig1t}. 
%
%
%
%

Thus, with the OLM, we found peaks  of CRS  in the photon number fluctuation spectra of SR LEDs. It supplements the results of \ct{Andre:19}, where CRS peaks  were predicted in the field  spectra. CRS peaks in the photon number fluctuation spectra are less visible than in the field spectra  (compare Figs.~\ref{Fig1t} and \ref{Fig2t}) and must be distinguished from the relaxation oscillation (RO) peaks. CRS and RO peaks are present in the field, and the photon number fluctuation spectra of SR lasers, see \ct{Protsenko_2021} and Figs.~\ref{Fig1t}, \ref{Fig2t} here. CRS peaks appear  at the SR LED regime. RO peaks appear  at high excitation when the laser generates coherent radiation. At the high excitation regime, ROs in the field spectra appear as tiny sidebands of the central peak, as in Fig.7a of \ct{Protsenko_2021}. The physical reason for the CRS peaks is the collective Rabi splitting. The physical reason for the RO peaks is the energy exchange between the lasing mode and the active medium.   

Now we analyze population fluctuations. We see in Eq.~\rf{MBE_St3} two sources of population fluctuations. The first source is fluctuations due to the pump and the energy decay  described by the two last terms on the right in Eq.~\rf{MBE_St3}. This source  
does not depend on the field and the active medium polarization, so 
it is the same in lasers with and without SR. 

The second source is fluctuations of the field-polarization interaction energy $\hat{\Sigma}$, which is  described by the term $\sim\delta\hat{\Sigma}$ on the right in Eq.~\rf{MBE_St3}.
As we see in Eq.~\rf{binary_eqs_fl_2}, $\delta\hat{\Sigma}$   depends on fluctuations $\delta\hat{D}$ of the dipole-dipole interaction energy of emitters. $\delta\hat{D}$ is large in SR lasers, so the contribution of $\delta\hat{\Sigma}$ to the population fluctuations is more significant in SR lasers than in lasers without SR. 

Let us see how the contributions of the field-polarization interaction energy and the pump-decay processes to population fluctuations depend on the laser parameters.
%
%
\begin{figure}

\bc\includegraphics[width=10cm]{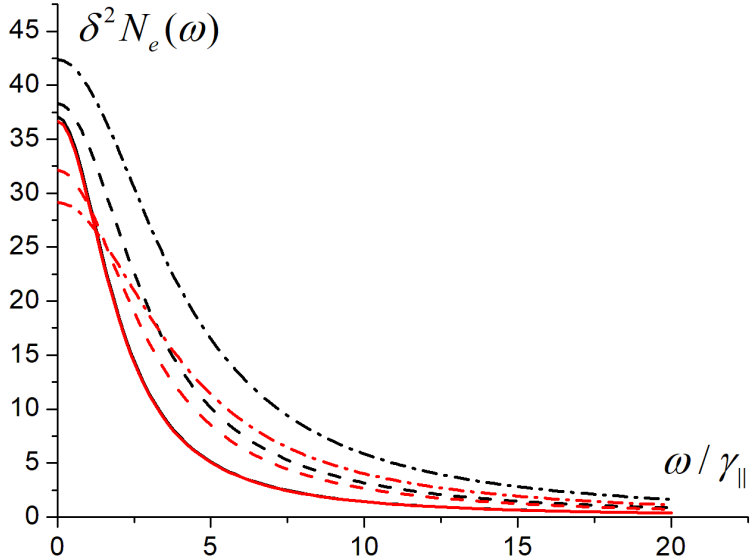}\ec
\caption{Population fluctuation spectra for the SR laser with $2\kappa/\gamma_{\perp} = 2>1$ (black curves) and the laser without SR with $2\kappa/\gamma_{\perp} = 1/30\ll 1$ (red curves). The normalised pump rate $P=1$ (solid curves), $P=2$ (dashed curves) and $P=3$ (dashed-dotted curves). When $P>1$ (dashed and dashed-dotted curves), population fluctuations in the SR laser  (black curves) are larger than in the laser without SR  (red curves). For small $P\leq 1$, population fluctuations depend mostly on the pump and the energy decay,  the same for lasers with and without SR, so the population fluctuation spectra for lasers with and without SR for $P\leq 1$ (solid red and black curves) practically coincide.}
\label{Fig3t}
\end{figure}
%
%
%
Fig.~\ref{Fig3t} shows population fluctuation spectra $\delta^2{\hat{N}}_e(\omega)$, given by Eq.~\rf{Pop_fl_sp}, for small $2\kappa/\gamma_{\perp} = 1/30\ll 1$ (the laser without SR, the red curves), and  large $2\kappa/\gamma_{\perp} = 2$ (the SR laser, the black curves), for various pump rates $P=1$ (solid curves) $P=2$ (dashed curves) and $P=3$ (dashed-dotted curves). 

Fig.~\ref{Fig3t} shows that the population fluctuation spectra for $P= 1$ and different $\gamma_{\perp}$  practically coincide. The same is true for $P<1$. The reason for it is  the contribution of the pump-decay fluctuations to population fluctuations, the same for lasers with and without SR,  which dominates at small pump $P\leq 1$.   

For $P>1$, population fluctuations in the SR laser   with  $2\kappa/\gamma_{\perp} = 2$   became  larger than in the laser without SR with small $2\kappa/\gamma_{\perp} = 1/30$. The difference in population fluctuations in SR lasers and lasers without SR grows with the pump $P$; see in Fig.~\ref{Fig3t} the difference between the dashed and the dotted-dashed curves of different colours. For $P>1$, population fluctuations depend more on the field and polarization interaction energy fluctuations. So population fluctuations became larger for the SR laser (the black dashed and dashed-dotted curves) than for the laser without SR (the red dashed and dashed-dotted  curves).

With a closer look at Fig.~\ref{Fig3t}, we see that population fluctuations in lasers without SR (red curves) are progressively (and nonlinearly) reduced with the pump. This effect is caused by well-known "population clamping" \ct{RevModPhys.68.127,PhysRevA.47.1431}  when the population inversion approaches $N_{th}$ and population fluctuations are reduced with the pump. Reduction of population fluctuations with the pump  leads to a well-known reduction of the laser linewidth with a factor of $1/2$. See more details, for example, in \ct{Protsenko_2021}.

\begin{figure}
\bc\includegraphics[width=10cm]{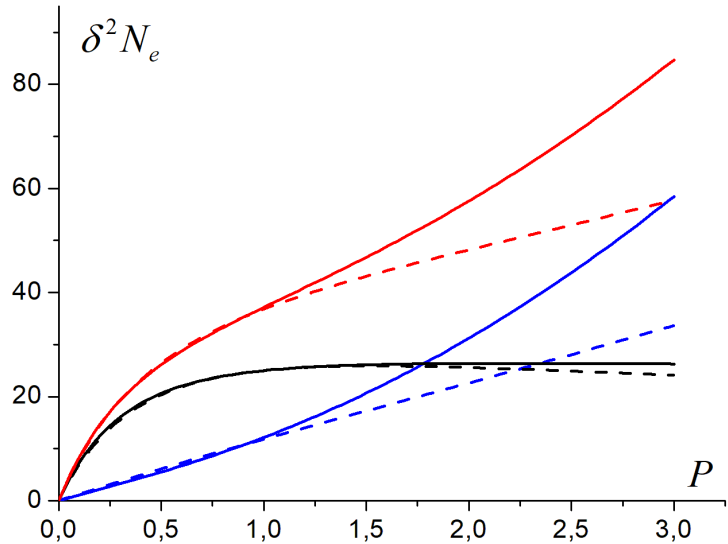}\ec
\caption{Population fluctuation dispersion is shown for the SR laser (the solid red  curve) and  the laser without SR (the dashed red curve). Black curves show contributions  due to the pump-decay fluctuations, and blue curves are  related to the field-polarisation interaction energy  fluctuations. Solid curves are for the SR laser; dashed curves are for the laser without SR. The pump-decay fluctuation contribution (black curves) dominates  at small pump $P<1$ and is practically the same for lasers with and without SR. The field-polarisation interaction energy contribution (blue curves) dominates at large pump $P>2$, which is more prominent for SR lasers than for lasers without SR.}
\label{Fig4t}
\end{figure}

Black curves in Fig.~\rf{Fig3t} show an opposite trend: population fluctuations in SR lasers grow with the pump. We explain this below, together with comments for Fig.~\ref{Fig4t}.

Fig.~\ref{Fig4t} shows the population fluctuation dispersion 
\beq
\delta^2N_e(P) = \frac{1}{2\pi}\int_{-\infty}^{\infty}\delta^2N_e(\omega)d\omega \lb{pop_disp}
\eeq
for $2\kappa/\gamma_{\perp} = 2$ (SR laser, solid curves) and for $2\kappa/\gamma_{\perp} = 1/30$ (laser without SR, dashed curves) and  contributions of the pump-decay and the field-polarisation interaction energy fluctuations into $\delta^2N_e(P)$. The red curves are $\delta^2N_e(P)$, the blue curves are parts of $\delta^2N_e$ related to the field and the polarization interaction energy fluctuations, and the black curves are parts of $\delta^2N_e$ related to the pump-decay fluctuations. 
%
%
%

For small pump $P<1$, the pump-decay fluctuations significantly affect  population fluctuations. They do not depend on $\gamma_{\perp}$ and are the same for lasers with and without SR; the red solid and dashed curves in Fig.~\ref{Fig4t} coincide. This effect has been used to simplify population fluctuation calculations at small pumps  \ct{Protsenko_2022}.  
For  $P>1$, the field-polarization interaction energy's influence on population fluctuations increases, corresponding part of population fluctuations  (blue curves in Fig.~\ref{Fig4t}) grew. 
This part is more significant for SR lasers than those without SR because of the dipole-dipole interaction energy $\hat{D}$ contribution to the field-polarisation interaction energy $\hat{\Sigma}$, see Eqs.~\rf{binary_eqs0_2}, \rf{binary_eqs_fl_2}. $\hat{D}$ and fluctuations $\delta\hat{D}$ are larger for SR lasers  than for lasers without SR. So, high dipole-dipole interaction energy fluctuations finally lead to significant population fluctuations in the SR lasers at a large pump.  

The contribution of fluctuations of  $\hat{D}$ into population fluctuations of SR lasers grows with the pump. It explains the increase of population fluctuations in SR lasers with the pump, as shown by black curves in Fig.~\ref{Fig3t}. Such population fluctuation increase   may be suppressed by population clamping at the higher pump. We leave the investigation of this for the future.

Let us discuss the OLM. The exact laser model is, of course, not an ensemble  of oscillators. However, OLM is a reasonable  approximation for the exact model in cases when population fluctuations can be neglected. OLM correctly predicts some mean values, such as the mean photon number, also for SR lasers  \ct{Protsenko_2021}, and leads to new results, such as the collective Rabi splitting in the field spectra of SR LEDs   \ct{Andre:19}. 

Formal justification of the OLM is the coincidence of equations, followed from of OLM, and  equations used previously in \ct{Protsenko_2021, Protsenko_2022,Andre:19}. We want to provide also physical arguments in favour of  OLM, following the interpretation of quantum inverted oscillators given in \ct{Stenholm_1986}. We quote \ct{Stenholm_1986}: "{\em A fully excited assembly of two-level atoms displays a ladder of equally spaced levels, which extends downwards for many steps. As long as the process is not saturating these levels behave exactly as an inverted oscillator...}". 

The present paper considers two assemblies of $N_e$ excited and $N_g$ emitters in the ground states.
There is evidence that the number of emitters in nanolasers is at least an order of magnitude larger than the number of photons at the threshold \ct{Kreinberg2017,doi:10.1063/1.5022958,PhysRevLett.126.063902}.
So we suppose  the mean number $n$ of the cavity photons is small, $n\ll N_{e,g}$ at a low laser excitation. In the OLM, we take a ladder of equally spaced levels, combined from the assembly of $N_e$ excited atoms, and describe it as an inverted oscillator with the operator $\hat{b}$ in the Hamiltonian \rf{Ham10}. This oscillator is hardly saturated (downward the energy levels in Fig.~\ref{Fig0t}) by a weak cavity field with the mean number of photons  $n \ll N_e$. Similarly, a non-inverted oscillator with the operator $\hat{c}$  combined from the assembly of $N_g$ atoms in the ground states can not be saturated by a weak field with  $n \ll N_g$. So with a weak lasing field and neglecting  fluctuations of populations of the lasing states, OLM is a physically reasonable approximation for the exact two-level laser model.

Let us consider some single emitter. In the stationary case, transitions between its levels are  population fluctuations, which we neglect. So, in our approximation, the emitter makes only virtual transitions. There are precisely $N_g$ emitters in the ground and $N_e$ -- in the excited states with no exchange  between manifolds of emitters in that states. When an emitter does not come into the excited (or de-excited) state, it does not matter whether it is a two-level system or an oscillator, so  we can replace emitters with oscillators in our approximation.  Polarization, non-zero at virtual transitions,  can be found separately for manifolds of the normal and the  inverted oscillators.  We do not eliminate polarization adiabatically in a difference with the quantum theory of linear amplifiers \ct{Stenholm_1986}.    

A harmonic oscillator is a primary system in the oscillation theory used in various areas of physics \ct{ANDRONOV1966583}. Many systems can be modelled as  interacting harmonic oscillators  in mechanics \ct{ANDRONOV1966583,Landau1976Mechanics}, optics  \ct{TKACHENKO200615},  chemistry \ct{OZEROV2007105} and other physical disciplines. We suggest that OLM  will be helpful for the analytical modelling  of lasers and other quantum optical systems with an active resonant medium. 
\section{\label{Sec5} Conclusion}
We present the oscillator laser model -- a quantum model of the two-level laser  in harmonic oscillators, including  {\em inverted} harmonic oscillators. OLM can be used when   population fluctuations of the lasing transitions can be neglected. OLM  is a zero-order approximation in the analytical approach of \ct{Protsenko_2021,Protsenko_2022,PhysRevA.59.1667,Andre:19}.   
With OLM, we calculate  zero-order diffusion coefficients \rf{dif_n0} for approximate equations \rf{binary_eqs_fl_1} -- \rf{binary_eqs_fl_3} for fluctuations in the photon number, the field-polarisation and the dipole-dipole interaction energies of the lasing medium.  Diffusion coefficients \rf{dif_n0} differ from the well-known ones obtained from generalized Einstein relations for the exact two-level laser model with Hamiltonian \rf{las_H}. Diffusion coefficients \rf{dif_n0} provide consistent results, in particular, for the photon number fluctuation variance \rf{ph_n_var}, found from  approximate equations \rf{MBE_St01}, \rf{MBE_St02} and \rf{binary_eqs_fl_1} -- \rf{binary_eqs_fl_3}. Zero-order diffusion coefficients are necessary to complete the approximate analytical procedure of solving laser equations \ct{Protsenko_2022}.

Here with  OLM, we calculate the photon number (or the intensity) fluctuation spectra of the lasing field and find the collective Rabi splitting peaks in the intensity fluctuation spectra of the superradiant lasers. This calculation supplements the results of \ct{Andre:19}, where the collective Rabi splitting peaks have been found in the lasing field spectra.
  
With the OLM, we calculate the population fluctuations in the first-order approximation, investigate and compare them in the SR lasers and the lasers without SR and suggest the mechanism for the growth of population fluctuations in SR lasers. Population fluctuations at a low laser excitation depend mainly on the pump-decay processes, the same in the lasers with and without SR.  When the laser excitation increases, the contribution from the field-polarisation interaction energy to the population fluctuations grows and overcomes the contribution of the pump-decay processes.  The dipole-dipole interaction between emitters, large in SR lasers, contribute to the field-polarisation interaction energy. So, when the laser excitation grows, the population fluctuations  become larger in SR lasers than in those without SR. 

Presented results show that population fluctuations play an essential role in SR lasers, as it is for SR \ct{PhysRevA.13.357}. We hope our results will be helpful for experimental studies of population fluctuations in SR lasers, for example, in possible experiments on observing and investigating parameters of the CRS peaks in the lasing field and the field intensity fluctuation spectra.
\section*{Appendix}
We solve the set of equations \rf{binary_eqs_fl_1} -- \rf{binary_eqs_fl_3} by Fourier-transform and find
%
\beqr
\delta\hat{n}(\omega )&=&\frac{-{{{\hat{F}}}_{n}}\left[ \left( i\omega -\kappa -{{\gamma }_{\bot }}/2 \right)\left( i\omega -{{\gamma }_{\bot }} \right)-2{{\Omega }^{2}}fN \right]+{{{\hat{F}}}_{\Sigma }}\Omega \left( i\omega -{{\gamma }_{\bot }} \right)-{{{\hat{F}}}_{D}}2{{\Omega }^{2}}f}{\left( i\omega -\kappa -{{\gamma }_{\bot }}/2 \right)\left[ 2\kappa {{\gamma }_{\bot }}-{{\omega }^{2}}-4{{\Omega }^{2}}fN-i\omega \left( 2\kappa +{{\gamma }_{\bot }} \right) \right]} \lb{sol_fc_1}\\
\delta \hat{\Sigma }(\omega )&=&\frac{{{{\hat{F}}}_{n}}2\Omega fN\left( i\omega -{{\gamma }_{\bot }} \right)-{{{\hat{F}}}_{\Sigma }}\left( i\omega -2\kappa  \right)\left( i\omega -{{\gamma }_{\bot }} \right)+{{{\hat{F}}}_{D}}\left( i\omega -2\kappa  \right)2\Omega f}{\left( i\omega -\kappa -{{\gamma }_{\bot }}/2 \right)\left[ 2\kappa {{\gamma }_{\bot }}-{{\omega }^{2}}-4{{\Omega }^{2}}fN-i\omega \left( 2\kappa +{{\gamma }_{\bot }} \right) \right]}. \lb{sol_fc_2}
\eeqr
%
Using diffusion coefficients \rf{dif_n0}, we calculate $\left<\delta \hat{n}(\omega)\delta \hat{n}(\omega')\right>$, $\left<\delta \hat{\Sigma}(\omega)\delta \hat{\Sigma}(\omega')\right>$, find the photon number fluctuation spectrum $\delta^2n(\omega)$ in 
$\left<\delta \hat{n}(\omega)\delta \hat{n}(\omega')\right>=\delta^2n(\omega)\delta (\omega+\omega')$,
\beqr
  \delta^2 n(\omega )S(\omega ) & = & 2\kappa n\left\{ \left[ 0.5\gamma_{\bot}^2-\omega^2+ \kappa\gamma_{\bot }\left(1-N/N_{th}\right) \right]^2+{{\omega }^{2}}{{\left( \kappa +3{{\gamma }_{\bot }}/2 \right)}^{2}} \right\}+
  \lb{ph_fl_sp}\\ 
 & & \left( \kappa {{\gamma }_{\bot }}/2{{N}_{th}} \right)\left[ 2\kappa (D+N_e)+{{\gamma }_{\bot }}({{N}_{0}}n+N_e)\right]\left( {{\omega }^{2}}+\gamma _{\bot }^{2} \right)+ \nonumber\\
 & &{{\kappa }^{2}}\gamma _{\bot }^{3}\left( N_0D+2{{N}_{e}}{{N}_{g}} \right)/N_{th}^{2}+4{{\kappa }^{2}}n\left[ \left( \kappa +{{\gamma }_{\bot }}/2 \right)\left( {{\omega }^{2}}+\gamma _{\bot }^{2} \right)-\kappa \gamma _{\bot }^{2}N/{{N}_{th}} \right]+\nonumber\\& & 2{{\kappa }^{2}}\gamma _{\bot }^{3}{{N}_{0}}/{{N}_{th}} \nonumber
\eeqr
and the spectrum $\delta ^{2}\Sigma(\omega)$ in $\left<\delta \hat{\Sigma}(\omega)\delta \hat{\Sigma}(\omega')\right>=\delta^2\Sigma(\omega)\delta (\omega+\omega')$, it is ${{\Omega }^{2}}{{\delta }^{2}\Sigma}(\omega) = S_{\Sigma}(\omega )/S(\omega )$ with
\beqr
 S_{\Sigma}(\omega ) &=& 
 2{{\kappa }^{3}}{{\gamma }_{\bot }}(nN/{{N}_{th}})\left( \gamma _{\bot }^{{}}nN/N_{th}^{{}}+4\kappa  \right)\left( {{\omega }^{2}}+\gamma _{\bot }^{2} \right)+ \lb{delta_Sigma}\\ 
 & &\left( \kappa {{\gamma }_{\bot }}/2{{N}_{th}} \right)\left[ 2\kappa D+{{\gamma }_{\bot }}{{N}_{0}}n+\left( 2\kappa +{{\gamma }_{\bot }} \right){{N}_{e}} \right]\left( {{\omega }^{2}}+4{{\kappa }^{2}} \right)\left( {{\omega }^{2}}+\gamma _{\bot }^{2} \right) \nonumber\\ 
 & &{{\kappa }^{2}}\gamma _{\bot }^{3}\left( {{\omega }^{2}}+4{{\kappa }^{2}} \right)\left[ \left( {{N}_{0}}D+2{{N}_{e}}{{N}_{g}} \right)/N_{th}^{2}+2n{{N}_{0}}/{{N}_{th}} \right] \nonumber
\eeqr
and
\beq
S(\omega )=\left[ {{\omega }^{2}}+{{\left( \kappa +{{\gamma }_{\bot }}/2 \right)}^{2}} \right]\left\{ {{\left[ 2\kappa {{\gamma }_{\bot }}(1-N/{{N}_{th}})-{{\omega }^{2}} \right]}^{2}}+{{\omega }^{2}}{{\left( 2\kappa +{{\gamma }_{\bot }} \right)}^{2}} \right\}. \lb{S_expr}
\eeq
 
\medskip

\bibliographystyle{MSP}
\bibliography{myrefs}

\end{document}